\documentstyle[eqsecnum,aps]{revtex}

\newcommand{\G}{G_{\mu \nu}}
\newcommand{\g}{g_{\mu \nu}}
\newcommand{\gu}{g^{\mu \nu}}
\newcommand{\gi}{g^{\alpha \beta}}
\newcommand{\tg}{\tilde{g}_{\mu \nu}}
\newcommand{\F}{\Phi}
\newcommand{\tG}{\tilde{G}_{\mu \nu}}
\newcommand{\tgu}{\tilde{g}^{\mu \nu}}
\newcommand{\tgi}{\tilde{g}^{\alpha \beta}}
\newcommand{\tC}{\tilde{\Lambda}(\F)}

\newcommand{\tT}{\tilde{T}}

\begin{document}

\draft

\title{ On the Energy-Momentum Tensor of the Scalar Field in Scalar--Tensor 
Theories of Gravity}

\author{ David I. Santiago\thanks{david@spacetime.stanford.edu}}
\address{ Department of Physics, Stanford University }
\author{ Alexander S. Silbergleit\thanks{gleit@relgyro.stanford.edu}} 
\address{ Gravity Probe B, W. W. Hansen Experimental Physics Laboratory, 
Stanford University}

\date{\today}

\maketitle

\begin{abstract}
We study the dynamical description of gravity, the appropriate definition of 
the scalar field energy-momentum tensor, and the interrelation between them 
in scalar-tensor theories of gravity. We show that the quantity 
which one would naively identify as the energy-momentum tensor of the scalar 
field is not appropriate because it is spoiled by a part of the dynamical
description of gravity. A new connection can be defined in terms of which the
full dynamical description of gravity is explicit, and the correct scalar field
energy-momentum tensor can be immediately identified. Certain inequalities must
be imposed on the two free functions (the coupling function and the potential) 
that define a particular scalar-tensor theory, to ensure that the scalar field 
energy density never becomes negative. The correct dynamical description leads 
naturally to the Einstein frame formulation of scalar-tensor gravity which is 
also studied in detail. 
\end{abstract}


\section{Introduction}

Scalar-tensor (ST) theories of gravity have an extensive history 
\cite{jr,fr,bd,brg,ndt,wag1,dm-ef}. Their study is important for a number of 
reasons, in particular because they provide the simplest generalization of 
Einstein's Theory of General Relativity (GR); they also turn out to be the low 
energy limit of certain attempts at a quantum theory of gravity, such as 
superstrings \cite{ss}.

In the present paper we review ST gravity in the physical or Jordan frame and 
the conformally related Einstein frame. 

The formulation of ST theory in the physical frame (sec. II) apparently does 
not lead to a well defined energy-momentum tensor for the scalar field. The 
scalar field terms on the right of the Einstein field equation, the quantities 
that one would naively associate with the energy-momentum tensor of the scalar 
field, prove to be ill behaved to constitute such a tensor: in the first place,
the scalar field energy density cannot be made universally nonnegative, as the 
presence of second derivatives makes it impossible. This has led some authors 
(see \cite{far,mag} and references therein) to reject the physical frame on 
exactly the grounds of the undesirable features of this ``apparent'' 
energy-momentum tensor. However, a careful look at it allows us to conclude 
that {\it all} scalar field terms on the right hand side of the Einstein 
equation  {\it should not be identified with the energy-momentum tensor of the 
scalar field}. In fact, the terms with the second covariant derivatives of the 
scalar field contain the connection, and hence a part of the dynamical 
description of gravity. This assertion is substantiated by the origin of the 
second derivative terms: they come from variation of the gravitational part of 
the action with respect to the metric. 

We find a new connection that describes the correct dynamics of gravity in sec 
III. The description in terms of this new connection removes the gravitational
dynamical terms from the right of the Einstein equation, leaving us with the 
correct energy-momentum tensor for the scalar field, the one that is not 
polluted by gravitational dynamical terms. The scalar field energy density can 
now be made nonnegative, which condition is implemented in the form of two 
important inequalities for the two otherwise arbitrary functions that define a 
particular ST theory (the coupling function and the potential, i.e., the scalar
field dependent cosmological "constant"). Moreover, the new connection arises 
from a metric conformally related to the physical metric. The conformal frame 
of this new metric is the Einstein frame. For completeness we develop ST 
gravity in the Einstein frame in sections IV and V of the paper.

Our main motivation for obtaining the described results  was a serious 
confusion in and long discussion of the meaning  and role of the Jordan and 
Einstein frames. For a good review of the different points of view and a very 
thorough list of references we refer readers to the article by Faraoni, 
\emph{et. al.} \cite{far}. The confusion is about what frame should be 
considered the "true physical" one. The source of the confusion is the fact 
that the Einstein frame metric is the one that correctly describes the spin-2 
dynamics of gravity. For this reason some authors single out the Einstein 
frame as the only one fundamental. On the other hand, by construction the 
metric in the Jordan frame is the one that determines metrical relations in 
spacetime, and particles move on geodesics of the physical metric, so the 
Jordan frame is claimed to be the fundamental one on these grounds. We think 
that no such recognition is pertinent, since it depends on whether one ascribes
more importance to the dynamics or to metrical relations and geodesics. 

There is even a belief that the two frames lead to different physics (see 
\cite{far,mag} and references therein). In our opinion, this statement should 
be taken only to the extent that the physical and Einstein frame metrics 
behave differently. Apart from that, those are just two different descriptions 
of the same physics. Physical properties of the continuum are determined by the
physical frame metric because this is the metric to which nongravitational 
fields couple universally by construction. Otherwise, since there is a well 
defined transformation between the two conformally related frames, one can work
in whatever frame is convenient as long as one uses the physical metric in the 
end to describe the physics of a particular problem.

The authors follow the basic conventions of Misner, Thorne, and Wheeler 
\cite{mtw} throughout the paper.

\label{sec-int}

\section{Scalar-Tensor Theories of Gravity in the Physical Frame}

We consider the most general scalar-tensor (ST) theories of gravity with a 
single scalar field. In these theories the gravitational interaction is 
mediated by  the metric $\tg$ and a spin-0 field, a scalar field $\F$.

The field equations for these theories follow from the action
\cite{jr,fr,bd,brg,ndt,wag1,dm-ef,dm-nd}
\begin{eqnarray}
S=\frac{1}{16\pi}\int d^{4}x\sqrt{-\tilde{g}}\,[\F \tilde{R} - 
\frac{\omega(\F)}{\F}\tgu\F_{, \mu}\F_{, \nu} -2\tC] \nonumber \\
 +\;\; S_{m}[\Psi_{m},\tg] \label{eq:aph} ,
\end{eqnarray}
where $ \ _{,\mu}$ represents the partial derivative with respect to $x^{\mu}$,
$\tilde{R}$ is the Ricci scalar constructed from the metric $\tg$, 
$\omega(\F)$ is the coupling function of the scalar field to matter, and the 
cosmological term $\tC$ is the scalar field potential. The scalar field $\F$ 
plays the role of the inverse gravitational constant $G^{-1}$. To ensure that 
gravity be attractive we impose the condition 
\begin{equation}
\F > 0
\end{equation}
The last term in (\ref{eq:aph}) is the action of the matter fields, $\Psi_{m}
$,  which couple only to the metric $\tg$ and not to the scalar field $\F$, in 
order to satisfy the weak equivalence principle. This formulation of ST gravity
is called the physical (or Jordan) frame description~\cite{dm-ef,dm-nd}, 
because the metric in this frame is the ``true'' metric of our spacetime. By 
``true'' we mean that this metric is the one measured by standard rods and 
clocks, i. e., it is the one that determines the geometry of our spacetime. 
The proper time measured by a moving test particle  is given by 
$d\tilde{\tau}^2 = \tg dx^\mu dx^\nu$. The 4-velocity, $\tilde{u}^\mu= dx^\mu 
/d\tilde{\tau}$, of the particle satisfies the geodesic equation
\begin{equation}
\tilde{u}^{\mu}\, _{, \beta}\tilde{u}^\beta +\tilde{\Gamma}^{\mu}_{\nu \beta}
\tilde{u}^\nu \tilde{u}^\beta=0 \label{eq:geo} \, ,
\end{equation}
where the connection $\tilde{\Gamma}^{\mu}_{\nu \beta}$ is the Christoffel 
symbol calculated with respect to the physical metric $\tg$.

So far we only assume that $\omega(\F)$ and $\tC$ are smooth enough functions 
on the positive semiaxis. The fact that $\omega(\F)$ could be negative might 
seem puzzling since the action (\ref{eq:aph}) would then appear to imply a 
negative kinetic term for the scalar field energy density. This is not the case
because in the physical frame it is not possible to define a suitable energy 
density for the scalar field due to its nonminimal coupling (through $\F 
\tilde{R}$) to the gravitational part of the action. Hence there is no suitable
geometrical definition of the energy-momentum tensor of the scalar field. The 
spin-2 and spin-0 excitations are ``entangled'' in the physical frame. In the 
following section we proceed to ``disentangle'' the two propagation modes by 
suitable transformations. Then it becomes possible to define the  
energy-momentum tensor and the energy density for the scalar field; the 
non-negativity of the latter requires  $\omega(\F)$ and $\tC$ to satisfy some 
inequalities. 

Each particular choice of the two arbitrary functions $\omega(\F)$ and $\tC$ 
specifies a different ST theory of gravity. In general, ST gravity represents 
theories with  cosmological and gravitational ``constants'' which change from 
point to point in spacetime.

We vary the action (\ref{eq:aph}) with respect to $\tg$ and $\F$ to obtain the 
field equations
\begin{eqnarray}
\tG=8\pi \frac{\tT_{\mu \nu}}{\F} +  \frac{\omega(\F)}{\F^2}( \F_{,\mu}\F_{,
\nu} - 1/2 \tg \tgi \F_{,\alpha} \F_{,\beta}) -\tg \frac{\tC}{\F} \nonumber \\
+\frac{1}{\F}(\tilde{\nabla_{\mu}} \tilde{\nabla_{\nu}} \F - \tg \Box_{\tilde{
g}} \F) \label{eq:tG}\\
\Box_{\tilde{g}}\F + \frac{1}{2}\F_{, \alpha} \F_{, \beta} \tgi \frac{d}{d\F} 
\ln \left(\frac{\omega(\F)}{\F} \right) + \frac{\F}{2 \omega (\F)} \left[ 
\tilde{R} - 2 \frac{d \tilde{\tC}}{d \F}\right] =0 \label{eq:F1}\, 
\end{eqnarray}
Here $\tilde{\nabla}_{\mu}$ is the covariant 
derivative with respect to $\tg$, $\Box_{\tilde{g}}\F= [\sqrt{-\tilde{g}}\tgu 
\F_{, \mu}]_{,\nu}/ \sqrt{-\tilde{g}}$ is the covariant D'Alambertian, $\tG
=\tilde{R}_{\mu\nu}-\left(1/2 \right)\tg \tilde{R}$ is the Einstein
tensor, and $\tilde{R}_{\mu\nu}$ is the Ricci tensor. By contracting equation 
(\ref{eq:tG}) and using the result to remove 
$\tilde{R}$ from the scalar field equation (\ref{eq:F1}) we find
\begin{equation}
\left[2 \omega (\F) + 3\right]\Box_{\tilde{g}}\F =  8 \pi \tT + 2 \F 
\frac{d\tC}{d \F} - 4 \tC - \frac{d \omega (\F)}{d \F} \F_{, \alpha} 
\F_{, \beta} \tgi, \label{eq:F}
\end{equation}
where $\tilde{T}_{\mu \nu}$ is the physical frame energy-momentum tensor of 
matter defined by the usual geometrical expression via the variational 
derivative as
\begin{equation}
\tilde{T}_{\mu \nu}=-\frac{2}{\sqrt{-\tilde{g}}}\frac{\delta S_{m}}{\delta 
\tgu} \, \label{eq:tT}
\end{equation}
Note that the values of the scalar field at which the coupling function 
$\omega (\F)$ turns to $-3/2$ are singular points of the scalar field equation 
(\ref{eq:F}).

Since $\tg$ is the metric which matter ``feels'', $\tT_{\mu \nu}$ is the 
``true'' energy-momentum tensor. Again, by ``true'' we mean that physically 
measurable quantities are the ones related to it. For example, an observer with
the 4-velocity $\tilde{u}^\mu$  would measure the energy density $\epsilon = 
\tilde{T}_{\mu \nu} \tilde{u}^\mu \tilde{u}^\nu$. Since the weak equivalence 
principle is satisfied (the matter fields only couple to $\tg$), the 
energy-momentum tensor of the matter fields is conserved:
\begin{equation}
\tilde{\nabla}_{\nu} \tilde{T}^{\mu \nu}=0. \label{eq:pcons}
\end{equation}
The direct derivation of this conservation law from equations (\ref{eq:tG}) and
(\ref{eq:F1}) is not at all straightforward.

\section{Energy-Momentum Tensor of the Scalar Field}

Let us now take a careful look at the Einstein field equation (\ref{eq:tG}). 
One usually associates the quantities on the right hand side with the 
energy-momentum tensor of the matter and non-gravitational physical fields. As 
mentioned above, $\tT_{\mu \nu}$ is the energy-momentum tensor of the matter 
fields as it comes from the variation of the matter action with respect to 
$\tg$. One is tempted to identify the scalar field terms with the 
energy-momentum tensor of the scalar field, but the situation is not that 
simple. On the right of equation (\ref{eq:tG}) there are terms which depend on 
the scalar field $\F$ and its first derivatives, and terms which are linear in 
the second order covariant derivatives of $\F$. The former terms,
\begin{equation}
\frac{\omega(\F)}{\F^2}( \F_{,\mu}\F_{,\nu} - 1/2 \tg \tgi \F_{,\alpha} \F_{,
\beta}) -\tg \frac{\tC}{\F} \, ,
\label{eq:3.1}
\end{equation}
come from varying the purely scalar field parts of  the action with respect to
$\tg$, that is, from varying everything that contains $\F$, except $\F 
\tilde{R}$ which describes some of the dynamics of the gravitational field. 
Hence expression (\ref{eq:3.1}) can be identified with a part of the scalar 
field energy-momentum tensor of the scalar field. The remaining scalar field 
terms on the right of equation (\ref{eq:tG}), those that have second covariant 
derivatives in them,
$$
\frac{1}{\F}(\tilde{\nabla_{\mu}} \tilde{\nabla_{\nu}} \F - \tg \Box_{\tilde{
g}} \F)\, ,
$$
should not belong to the scalar field energy-momentum tensor. First of all, the
presence of second derivatives of $\F$ is undesirable because it would make it 
impossible to have a nonnegative energy density for the scalar field. Even 
more important, these terms contain a combination
$$
\frac{1}{\F}(\tg \tilde{\Gamma}^{\alpha}_{\sigma \alpha}\tilde{g}^{\sigma \beta
}\F_{\beta} - \tilde{\Gamma}^{\alpha}_{\mu \nu}\F_{\alpha}) \, 
$$
with the connection in it. It comes, after integration by parts, from varying 
$\F \tilde{R}$ which  carries the dynamical information on the tensor part of 
gravity. Hence these terms should be regarded as a part of the dynamical 
evolution and constraint equations for the gravitational field. Therefore 
these terms may belong on the left hand side of equation (\ref{eq:tG}) where 
the rest of the dynamical description of the gravitational field (in the form 
of the Einstein tensor $\tG$) resides. 

In order to have all the terms that contain the dynamical description of 
gravity on the left hand side of the Einstein equation (\ref{eq:tG}) we define 
a new connection
\begin{equation}
\Gamma^{\alpha}_{\beta \gamma} = \tilde{\Gamma}^{\alpha}_{\beta \gamma} +
\frac{1}{2\F} \left( \delta^{\alpha}_{\beta} \F_{, \gamma} + \delta^{\alpha}_{
\gamma} \F_{, \beta} - \tilde{g}_{\beta \gamma} \F^{, \alpha} \right) \, ,
\label{eq:con}
\end{equation}

($\F^{,\alpha} \equiv \tilde{g}^{\alpha \sigma} \F_{, \sigma}$). That 
$\Gamma^{\alpha}_{\beta \gamma}$ transforms as a connection follows from the
fact that
\begin{equation}
D^{\alpha}_{\beta \gamma} \equiv \frac{1}{2 \F}\left( \delta^{\alpha}_{\beta} 
\F_{, \gamma} +\delta^{\alpha}_{\gamma} \F_{, \beta} - \tilde{g}_{\beta \gamma}
\F^{, \alpha} \right) \label{eq:D}
\end{equation}
transforms as a $(1,2)$ tensor. Since
$\tilde{\Gamma}^{\alpha}_{\beta \gamma}$ is the connection compatible with the 
metric ($\tilde{\nabla}_{\alpha} \tg =0$), the connection $\Gamma^{\alpha}_{
\beta \gamma}$ is generally not compatible with the metric. Indeed, one easily 
finds $\nabla_{\alpha} \tg = - \tg \F_{,\alpha}/ \F$, where $\nabla$ represents
the covariant derivative with respect to the connection $\Gamma^{\alpha}_{\beta
\gamma}$. In the Appendix we derive the following relation $(R_{\mu \nu}$ is 
the Ricci tensor calculated from the connection $\Gamma^{\alpha}_{\beta \gamma}
$ ):
\begin{equation}
\tilde{R}_{\mu \nu} = R_{\mu \nu} + \frac{1}{\F} \left( \tilde{\nabla_{\mu}} 
\tilde{\nabla_{\nu}} \F + \frac{1}{2}\tg \Box_{\tilde{g}} \F \right) -
\frac{3}{2} \frac{\F_{,\mu} \F_{,\nu}}{\F^2} \, 
\label{eq:3.4}
\end{equation}
If we now define the Ricci scalar and Einstein tensor with respect to the new 
connection as
\begin{equation}
 R^* \equiv \tgu R_{\mu \nu}, \qquad
\G \equiv R_{\mu \nu} - \frac{1}{2} \tg R^* \, ,
\end{equation}
we then obtain:
\begin{eqnarray}
 \tilde{R} = R^* + \frac{3 \Box_{\tilde{g}} \F}{\F} - \frac{3}{2} \frac{\F^{,
\alpha} \F_{,\alpha}}{\F^2}\, , \label{eq:ricc}\\
\tG = \G +  \frac{1}{\F}(\tilde{\nabla_{\mu}} \tilde{\nabla_{\nu}} \F - \tg 
\Box_{\tilde{g}} \F) \nonumber \\  
- \frac{3}{2 \F^2} ( \F_{,\mu}\F_{,\nu} - 1/2 \tg \tgi \F_{,\alpha} \F_{,\beta}
) \label {eq:Eins}\, .
\end{eqnarray}
One immediately sees that the right hand side of the last equation includes the
``offending'' second covariant derivative terms, which will cancel when we 
substitute (\ref{eq:Eins}) in the Einstein equation (\ref{eq:tG}). In terms of 
the new Einstein tensor, $\G$, the latter becomes
\begin{equation}
\G=8\pi \frac{\tT_{\mu \nu}}{\F} +  \frac{\omega(\F) + 3/2}{\F^2}( \F_{,\mu}
\F_{,\nu} - 1/2 \tg \tgi \F_{,\alpha} \F_{,\beta}) -\tg \frac{\tC}{\F} 
 \label{eq:G0} \, .
\end{equation}
We see that we have succeeded in eliminating the terms containing the 
connection on the right hand side with an additional bonus that the full second
covariant derivatives of the scalar field were eliminated. The dynamical 
description of gravity is in terms of the connection $\Gamma^{\alpha}_{\beta 
\gamma}$ and not the metric connection $\tilde{\Gamma}^{\alpha}_{\beta \gamma}
$. Hence, $\G$ describes the complete dynamics of the gravitational field and, 
therefore, we can identify the quantities to its right in equation 
(\ref{eq:G0}) as the energy-momentum tensor of the matter and physical fields. 
We have effectively ``disentangled'' the tensor and scalar modes.

In particular the energy-momentum tensor, $\tilde{\Sigma}_{\mu \nu}$, of the 
scalar field is given by
\begin{equation}
8 \pi \frac{\tilde{\Sigma} _{\mu \nu}}{\F} \equiv \frac{\omega(\F) + 3/2}{\F^2}
( \F_{,\mu}\F_{,\nu} - 1/2 \tg \tgi \F_{,\alpha} \F_{,\beta}) -\tg 
\frac{\tC}{\F} \,;
\label{eq:Sig}
\end{equation}
note that unlike the energy-momentum tensor of matter, $\tT_{\mu \nu}$ , it is 
not, of course, covariantly conserved:
$$
\tilde{\nabla}_{\mu} \tilde{\Sigma}^{\mu}_{\nu} = \frac{\F_{,\nu}}{2\F} \left[
\tilde{T} - \frac{\tC}{2 \pi}  - \frac{\omega(\F) + 3/2}{8 \pi 
\F}\F_{,\alpha} \F^{,\alpha}\right]
\not\equiv0
$$
This nonconservation is a consequence of the nonminimal coupling between the 
scalar field and the metric.

We now impose the condition that the energy density $\tilde{\Sigma} _{\mu \nu}
\tilde{u}^\mu \tilde{u}^\nu$ of the scalar field for any observer be 
nonnegative. Expression (\ref{eq:Sig}) shows that this can be achieved if and 
only if 
\begin{eqnarray}
\omega(\F) \ge -\frac{3}{2} \label{eq:om}\\
\tC \ge 0 \label{eq:lam} \, ;
\end{eqnarray}
we consider these inequalities valid for our ST gravity. The classical 
treatment of the field equations requires the functions $\omega(\F)$ and $\tC$ 
to have at least one continuous derivative for $\Phi>0$, which condition we 
always assume in this paper.

Note that if we define a new metric, $\g$, by the conformal transformation $\g 
= \F \tg$, then the Christoffel symbol $\{ ^{\; \alpha}_{\beta \gamma} \}$ with
respect  to $\g$, i. e., the connection compatible with $\g$, satisfies
\begin{equation}
\left\{ ^{\; \alpha}_{\beta \gamma} \right\} = \tilde{\Gamma}^{\alpha}_{\beta 
\gamma} + \frac{1}{2\F} \left( \delta^{\alpha}_{\beta} \F_{, \gamma} + 
\delta^{\alpha}_{\gamma} \F_{, \beta} - \tilde{g}_{\beta \gamma} \F^{, \alpha} 
\right) \, .
\label{eq:3.12}
\end{equation}
By (\ref{eq:3.12}) and (\ref{eq:con}) we have $\Gamma^{\alpha}_{\beta \gamma} 
= \{ ^{\; \alpha}_{\beta \gamma} \}$; therefore
$\Gamma^{\alpha}_{\beta \gamma}$ is the metric connection for $\g$. We conclude
that in ST gravity there is a metric $\tg$ which determines proper lengths and
times, and geodesics in our spacetime, i. e., it is a metric in the proper 
sense of the word. There is also the conformally related ``metric'' $\g$ which 
carries the dynamical information of the gravitational field, in other words it
describes the pure spin-2 excitations. However, it is important to bear in mind
that the expression $\g dx^\mu dx^\nu$ does not represent a physical spacetime 
interval.

We call $\g$ the dynamical metric and 
$\Gamma^{\alpha}_{\beta \gamma}$ the dynamical connection. The corresponding 
frame conformally related to the physical frame is called the Einstein frame 
\cite{dm-ef,dm-nd,far}.

\section{Scalar-Tensor Theories of Gravity in the Einstein Conformal Frame}

We now proceed to  develop the description of ST gravity in the Einstein frame.
As we mentioned, this is the description in terms of the conformally related 
dynamical metric and connection. Instead of working from the field equations,
it is more convenient to work from the action (\ref{eq:aph}). The only 
dynamical geometrical quantity in the action is the Ricci scalar $\tilde{R}$.
In the previous section we derived an expression for this quantity in terms
of the Ricci scalar $R^*$ defined by contracting the dynamical Ricci scalar
$R_{\mu \nu}$ with the physical metric $\tgu$. The appropriate Ricci scalar for
the Einstein frame description is defined by contracting with the dynamical 
metric: 
\begin{equation}
R= \gu R_{\mu \nu}= \frac{\tgu}{\F} R_{\mu \nu}= \frac{R^*}{\F} \, .
\end{equation}
In terms of $R$, equation (\ref{eq:ricc}) becomes
\begin{equation}
\tilde{R} = \F R + \frac{3 \Box_{\tilde{g}} \F}{\F} - \frac{3}{2} \tilde{g}^{
\alpha \beta} \frac{\F_{,\alpha} \F_{,\beta}}{\F^2}\, . 
\end{equation}
The quantities on the right hand side have to be expressed in terms of the 
dynamical metric. Since $\g =  \F \tg$ we have
$$
\tgu = \F \gu,\qquad
\sqrt{-\tilde{g}}=\frac{\sqrt{-g}}{\F^2}. \qquad
\Box_{\tilde{g}} \F =\frac{1}{\sqrt{-\tilde{g}}} \left[ \sqrt{-\tilde{g}} \tgu 
\F_{,\mu}\right]_{,\nu} =
 \frac{\F^2}{\sqrt{-g}} \left[ \frac{\sqrt{-g}}{\F} \gu 
\F_{,\mu}\right]_{,\nu} = \F^2 \Box \ln \F \, .
$$
Using these relations we obtain
\begin{equation}
\tilde{R} = \F R + 3 \F \Box \ln \F - \frac{3}{2} \frac{\F^{,\alpha} \F_{, 
\alpha}}{\F}\, , \label{eq:ricci}
\end{equation}
where from now on all indices are raised and lowered with the dynamical metric 
$\g$, i. e. $\F^{, \mu} = \gu \F_{,\nu}$, unless otherwise indicated. The 
action (\ref{eq:aph}) in terms of the dynamical metric becomes
\begin{eqnarray}
S=\frac{1}{16\pi}\int d^{4}x\sqrt{-g}\,[ R - \frac{\omega(\F) + 3 /2 }{\F^2}
\gu\F_{, \mu}\F_{, \nu} -2\frac{\tC}{\F^2}] + S_{m}[\Psi_{m},\g / \F] 
\label{eq:a} \, ,
\end{eqnarray}
where the term containing $3 \Box \ln \F$ was integrated by parts to give zero 
by supposing all our fields vanish on the boundary of spacetime.

Let us pause briefly to mention a few properties of our Einstein frame action 
(\ref{eq:a}). One immediately notices that the metric and scalar field parts 
are now untangled in the sense that no scalar field dependent factor stands in 
front of $R$ in (\ref{eq:a}). The dynamics of gravity is described by the Ricci
scalar $R$, which now appears by itself without any scalar field dependent 
factors. There is also the scalar field part of the action in which the only 
coupling to gravity is through the metric; it leads thus to a well defined 
energy-momentum tensor for the scalar field. Moreover, it is clear that the two
conditions (\ref{eq:om}) and (\ref{eq:lam}) must be satisfied, to have a 
nonnegative scalar field energy density. On the other hand, the separation of 
the dynamics of the metric and scalar field comes with a price: matter couples 
to gravity nonminimally (but universally) through the physical metric $\tg = \g
 / \F$, to preserve the weak equivalence principle.

The Einstein frame field equations follow by varying the action 
(\ref{eq:a}). When it is varied with respect to the dynamical metric $\gu$, 
the Einstein field equation
\begin{equation}
\G=8\pi T_{\mu \nu} +  \frac{\omega(\F) + 3/2}{\F^2}( \F_{,\mu}
\F_{,\nu} - 1/2 \g \gi \F_{,\alpha} \F_{,\beta}) -\tg \frac{\tC}{\F^2}
\equiv 8\pi\left( T_{\mu \nu}+\Sigma_{\mu \nu}\right)
 \label{eq:G} 
\end{equation}
is obtained. The Einstein tensor on its left describes the evolution and 
constraints for the dynamical metric. In the right hand side are the 
sources, namely, the energy-momentum tensor of the scalar field defined in 
(\ref{eq:Sig}) and the Einstein frame energy-momentum tensor, $T_{\mu \nu}$, of
the matter fields whose standard definition is
\begin{equation}
T_{\mu \nu}=-\frac{2}{\sqrt{-g}}\frac{\delta S_{m}}{\delta \gu} \, . 
\label{eq:T}
\end{equation}
We stress that the Einstein frame energy-momentum tensor does not represent the
energy-momentum tensor of the matter fields, because it is defined by the 
variation with respect to the  dynamical metric which is not the metric of the 
physical continuum. As we mentioned earlier, the physical frame energy-momentum
tensor is the ``true'' one as it is the physical metric that defines metrical 
relations in spacetime. From the definitions of the energy-momentum tensors of 
matter in each frame ( see equations (\ref{eq:tT}) and (\ref{eq:T})) the 
relations between their components follow easily: 
\begin{equation}
T_{\mu \nu} = \frac{\tT_{\mu \nu}}{\F},\qquad
T^{\mu}_{\nu} = \frac{\tT^{\mu}_{\nu}}{\F^2},\qquad
T^{\mu \nu} = \frac{\tT^{\mu \nu}}{\F^3},\qquad
T \equiv T^{\mu}_{\mu}  =  \frac{\tT^{\mu}_{\mu}}{\F^2} \equiv \frac{\tT}{\F^2}
\end{equation}
Here indices for the Einstein frame tensor are raised and lowered with the 
dynamical metric $\g$ and for the physical frame tensor with the physical 
metric $\tg$. The conservation of the physical energy-momentum tensor 
(\ref{eq:pcons}) leads to the following differential equation for the Einstein 
frame tensor:
\begin{equation}
T^{\mu}_{\nu}\,_{; \mu}= - \frac{\F_{, \nu}}{2\F} T \, , \label{eq:cons}
\end{equation}
where $;$ represents covariant differentiation with respect to the dynamical
metric. The Einstein frame energy-momentum tensor is not 
covariantly conserved, hence free particles do not follow geodesics of the 
dynamical metric. 

The wave equation for the scalar field is obtained by varying the action
(\ref{eq:a}) with respect to $\F$:
\begin{equation}
\left[\omega(\F) +3/2\right]\left(\Box \F - \frac{\F_{, \alpha} \F^{, \alpha}}{
\F}\right) ={\F^2}\left\{ \frac{4 \pi T}{\F} + \frac{d}{d \F} \left[ \frac{
\tC}{\F^2} \right]\right\}-\frac{1}{2}\F_{, \alpha} \F^{, \alpha}\frac{d
\omega(\F)}{d\F} \label{eq:Phi2}
\end{equation}
Using this we demonstrate that the scalar field energy-momentum
tensor is not covariantly conserved:
\begin{equation}
\Sigma^{\mu}_{\nu}\,_{; \mu} = \frac{\F_{, \nu}}{2\F} T \, . \label{eq:Phicons}
\label{eq:S}
\end{equation}
However, from (\ref{eq:cons}) and (\ref{eq:S}) it is clear that the sum of the
matter and scalar field energy-momentum tensors is covariantly conserved as it 
must because of the contracted Bianchi identity, $G^{\mu}_{\nu}\,_{; \mu}=0$.

Note that in the Einstein frame we have an appropriate description of the 
tensor part of gravity and of the scalar field. Moreover the energy-momentum 
tensor for the scalar field can only be defined in terms of the dynamical 
connection and/or dynamical metric. On the other hand, it costs a more 
complicated description of matter, with a nonconserved energy-momentum tensor. 
In the physical frame the description of gravity and of the scalar field are 
complicated since their propagation modes are entangled, but the description of
matter is simple. Each frame has its technical advantages, but one must 
remember that it is the physical metric and energy-momentum tensor which are 
most directly related to observables.

\section{ Scalar Field Redefinition in the Einstein Frame}

As stated in the previous section, the Einstein frame provides the description 
of our gravitational theory in terms of the dynamical metric $\g= \F \tg$. One 
has yet a freedom to redefine the scalar field, which  choice is usually used 
to simplify the scalar field equation (\ref{eq:Phi2}), complicated so far by 
the presence of terms quadratic in the scalar field derivatives. To do that, 
one realizes that the latter originate from the term
\begin{equation}
\frac{\omega(\F) + 3 /2 }{\F^2} \gu\F_{, \mu}\F_{, \nu} \, .
\label{eq:5.1}
\end{equation}
in the action (\ref{eq:a}). If the factor in front of the derivative terms were
a constant, the scalar field equation would simplify significantly. In view of 
the inequality (\ref{eq:om}) imposed on the coupling function $\omega(\F)$, 
this immediately gives rise to the two following cases.
\medskip

Case 1. $\omega(\F)\equiv -3/2$

The coefficient in (\ref{eq:5.1}) is not just a constant, but exactly zero, so 
no field redefinition is needed. This is a rather peculiar situation from the 
physical standpoint, since both the energy momentum tensor of the scalar field 
(\ref{eq:Sig}) and the Einstein frame action (\ref{eq:a}) contain no kinetic 
terms at all. As a consequence the scalar field equation (\ref{eq:Phi2}) 
reduces to 
\begin{equation}
\frac{4 \pi T}{\F} + \frac{d}{d \F} \left[ \frac{\tC}{\F^2}  \right] =0 
\label{eq:5.2}
\end{equation}
Instead of being an evolution equation, (\ref{eq:5.2}) is rather an algebraic 
equation for the scalar field $\F$ which allows for its direct determination 
from the matter term by means of inversion. The physical plausibility of this 
case seems questionable to us. 
\medskip

Case 2. $\omega(\F)\not\equiv -3/2$

In this case we define a new scalar field $\chi=\chi(\F)$ by the equations ($\F_0>0$ is a constant) 
\begin{equation}
\sqrt{2} \frac{d \chi}{d \ln \F} = \sqrt{\omega(\F) + 3 /2},\qquad 
\chi(\F)=\frac{1}{\sqrt2}\int_{\F_0}^\F  \frac{\sqrt{\omega(\F) + 3 /2}}{\F}\,d\F,
\label{eq:defchi}
\end{equation}
so that the coefficient in (\ref{eq:5.1}) turns into $2$ after replacing $\F$ 
by $\chi$. 

It is clear that the new scalar field $\chi$ is a non-decreasing 
function of the old one, $\F\geq0$, and its growth can stop only at such values
$\F_*$ for which $\omega(\F_*)=-3/2$; if they exist, those are evidently the 
inflection points of $\chi(\F)$. Therefore an inverse to $\chi(\F)$ function $\F=\F(\chi)$ is uniquely 
determined; it is a monotonically increasing function of its argument.
 
The range of the new scalar field $\chi(\F)$ depends on the 
value of the coupling function at $\F=+0$. If  $\omega(+0)= -3/2$ and the integral in (\ref{eq:defchi}) converges at $\F=0$, then the range is
$$
-\infty<\chi(+0)<0\leq\chi(\F)\leq\chi(+\infty)\leq+\infty,\qquad 0\leq\F\leq+\infty,
$$
If, on the other hand, $-3/2<\omega(+0)\leq\infty$, then the integral in (\ref{eq:defchi}) 
diverges at $\F=0$, and the range includes the negative semiaxis:
$$
-\infty=\chi(+0)\leq\chi(\F)\leq\chi(+\infty)\leq+\infty,
\qquad 0\leq\F\leq+\infty,
$$
The upper limit $\chi(+\infty)$ of the range is either infinite or finite positive depending on whether the integral in (\ref{eq:defchi}) diverges or converges at $\F=+\infty$. For the second possibilty to occur, the coupling function must tend to $-3/2$ at infinity, so that the generic range of $\chi(\F)$, without any additional assumptions about $\omega(\F)$, is the whole real axis. 
\smallskip

After this redefinition of the scalar field the Einstein frame action and field
equations become
\begin{equation}
S=\frac{1}{16\pi}\int d^{4}x\sqrt{-g}\,[ R - 2 \gi \chi_{, \alpha} \chi_{,
\beta} -2\Lambda(\chi)] + S_{m}[\Psi_{m},\g /\F
(\chi)] 
\label{eq:as} 
\end{equation}
\begin{equation}
\G=8\pi T_{\mu \nu} + 2( \chi_{, \mu} \chi_{, \nu} - 1/2 \g \gi \chi_{, \alpha}
\chi_{, \beta}) -\g \Lambda(\chi) \label{eq:Gs}
\end{equation}
\begin{equation}
\Box \chi = \frac{1}{2} \frac{d\Lambda(\chi)}{d \chi}  +
\left[ 2 \pi  \frac{d \ln \F(\chi)}{d \chi}\right]T ,
\label{eq:chi}
\end{equation}
where the function $\Lambda(\chi)$ is defined as
\begin{equation}
\Lambda(\chi) = \tilde{\Lambda}(\F
(\chi))/\F^2(\chi),
\label{eq:5.9}
\end{equation}
and $T_{\mu \nu}$ is the Einstein frame energy-momentum tensor of matter 
(\ref{eq:T}) (its interpretation and relations to the physical energy-momentum 
tensor were discussed in the previous section). The scalar field equation 
(\ref{eq:chi}) has the same singular points as the equation (\ref{eq:F}), 
namely, the values of $\F$ at which the coupling function turns to $-3/2$, 
because $d \F(\chi)/d \chi=\infty$ at these points.

Note that in principle one can, of course, use the negative branch of the 
square root in the definition (\ref{eq:defchi}) of the function $\chi(\F)$, or 
even combine the positive and negative branches (to keep the derivative $d \chi
/d \F$ continuous, the change of the branches can only occur at those singular 
points, if any). However, since this is just a transformation function, one 
does not care for making it as general as possible. On the contrary, its 
simplest form is most valuable as soon as the goal of the transformation, that 
is the wave equation simplification, is achieved. Our choice of $\chi(\F)$ can 
always provide it non-decreasing, i. e., with no extrema whatsoever. Its 
derivative turns to zero at most at the inflection points where $\omega(\F)=-3
/2$, if such points exist.

We have thus transformed from a description in terms of the scalar field $\F$ 
and the two arbitrary functions $\omega(\F)$ and $\tC$, satisfying conditions 
(\ref{eq:om}) and (\ref{eq:lam}), respectively, 
to a description in terms of the scalar field 
$\chi$ and the two arbitrary functions $\F(\chi)$ and $\Lambda(\chi)$.  The 
scalar field equation (\ref{eq:chi}) is a wave equation with a potential $ 
\Lambda(\chi)$. The source for the scalar field equation is proportional to the
trace of the Einstein frame energy-momentum tensor. The factor of 
proportionality describes the coupling between matter and scalar field; whith 
our choice of transformation, this factor is positive for $0<\F<\infty$ (by 
(\ref{eq:defchi}), the derivative $d\ln\F/d\chi$ can only turn to zero when $\F
\rightarrow\+0$ or $\F\rightarrow\infty$, and only if $\omega(\F)$ tends to 
infinity in the corresponding limit). Generally, the factor is scalar field 
dependent; the special case when it does not depend on the scalar field 
corresponds to $\omega=const$, i. e., to the Brans-Dicke theory.

\section{Conclusions}

We constructed the proper energy-momentum tensor of the scalar field in 
scalar-tensor gravity, disentangling simultaneously the dynamics of the scalar 
field from that of gravity {\it per se}. The Einstein frame arises naturally 
out of this disentanglement .

We have shown that {\it all} scalar field terms on the right of the Einstein 
equation (\ref{eq:tG}) {\it cannot be identified} with the energy-momentum 
tensor of the scalar field because some of them contain the second covariant 
derivatives. The latter originate from variation of the gravitational part of 
the action, $\F \tilde{R}$, after an integration by parts. Hence they form  a 
part of the dynamical description of gravity, and not of the scalar field. They
occur because the dynamics of gravity and that of the purely scalar excitations
are entangled in the physical frame, as a result of the nonminimal coupling 
between gravity and the scalar field. We defined a new connection in terms of 
which the full dynamics of the gravitational part can be explicitly separated. 
When doing this, one immediately finds the correct energy-momentum tensor of 
the scalar field as given by equation (\ref{eq:Sig}). This is a well defined 
energy-momentum tensor as long as we impose the condition that the energy 
density never becomes negative. This condition leads to the inequalities 
(\ref{eq:om}) and (\ref{eq:lam}) for the two otherwise arbitrary functions 
specifying the theory, the coupling function and the scalar field dependent 
cosmological "constant". We regard these restrictions as necessary for any ST 
theory to be physical.

It is of importance that the dynamical connection can also be obtained by a 
conformal transformation of the metric. The conformally transformed metric is 
the geometrical object that describes the full dynamics of the tensor part of 
gravity. The Einstein conformal frame is the one defined in terms of the 
dynamical metric. Usually, the scalar field is redefined in the conformal 
frame, to simplify the scalar field equation in it. We analyze the behavior of 
the appropriate transformation function and demonstrate that it can always be 
chosen monotonic.

Our hope is that the results presented here shed some light on the subject of
the relationship between the physical and Einstein frames, which seems to have 
caused not a small amount of confusion before. In particular, the behavior of 
different quantities in each frame,  such as the scalar field and its 
energy-momentum tensor, becomes clear, as well as the relation between the 
physical and dynamical metrics, and the correct dynamical description of 
gravity in scalar-tensor theories.

\section*{Acknowledgments}

This work was supported by NASA grant NAS 8-39225 to Gravity Probe~B. We are 
grateful to R.V.Wagoner for many valuable comments and to the Gravity Probe B 
Theory Group for fruitful discussions.

\appendix
\section{Relations between physical and dynamical Riemann and Ricci Tensors}

Here we derive the relations between the Riemann and Ricci 
tensors constructed from  the physical and dynamical connections.
The physical Riemann tensor is defined by
\begin{equation}
\tilde{R}^{\alpha} \:_{\beta \gamma \delta} = \tilde{\Gamma}^{\alpha}_{\beta 
\delta , \gamma} - \tilde{\Gamma}^{\alpha}_{\beta \gamma, \delta} +  
\tilde{\Gamma}^{\alpha}_{\mu \gamma} \tilde{\Gamma}^{\mu}_{\beta \delta} -
\tilde{\Gamma}^{\alpha}_{\mu \delta} \tilde{\Gamma}^{\mu}_{\beta \gamma} 
\label{eq:1}
\end{equation}
The dynamical Riemann tensor, $R^{\alpha} \:_{\beta \gamma \delta}$, is defined
similarly with the dynamical connection, $\Gamma^{\alpha}_{\beta \gamma}$, in place of the physical one, $\tilde{\Gamma}^{\alpha}_{\beta \gamma}$.

Defining the one-form
\begin{equation}
A_\alpha \equiv (\ln \F)_{, \alpha} = \frac{\F _{, \alpha}}{\F} \label{eq:A}
\end{equation}
equation (\ref{eq:D}) becomes
\begin{equation}
D^{\alpha}_{\beta \gamma} = \frac{1}{2}(\delta^{\alpha}_{\beta} A_\gamma +
\delta^{\alpha}_{\gamma} A_\beta - \tilde{g}_{\beta \gamma} A^\alpha) ,
 \label{eq:D2}
\end{equation}
and equation (\ref{eq:con}) relating the two connections is
$$
\Gamma^{\alpha}_{\beta \gamma} = \tilde{\Gamma}^{\alpha}_{\beta \gamma} +
D^{\alpha}_{\beta \gamma}
\label{eq:con2}
$$
Using this in the equation (\ref{eq:1}) we obtain the following relation between the two
Riemann tensors:
$$
R^{\alpha} \:_{\beta \gamma \delta} = \tilde{R}^{\alpha} \:_{\beta \gamma 
\delta} + D^{\alpha}_{\beta \delta , \gamma} - D^{\alpha}_{\beta \gamma, 
\delta} + D^{\alpha}_{\mu \gamma} \tilde{\Gamma}^{\mu}_{\beta \delta} +
\tilde{\Gamma}^{\alpha}_{\mu \gamma} D^{\mu}_{\beta \delta} \nonumber \\
-D^{\alpha}_{\mu \delta} \tilde{\Gamma}^{\mu}_{\beta \gamma} - \tilde{\Gamma}^{
\alpha}_{\mu \delta} D^{\mu}_{\beta \gamma} + D^{\alpha}_{\mu \gamma} D^{\mu}_{
\beta \delta} - D^{\alpha}_{\mu \delta} D^{\mu}_{\beta \gamma}
$$
With some patience one verifies that
$$
\tilde{\nabla}_{\gamma} D^{\alpha}_{\beta \delta} - \tilde{\nabla}_{\delta} 
D^{\alpha}_{\beta \gamma} = D^{\alpha}_{\beta \delta , \gamma} - D^{\alpha}_{
\beta \gamma, \delta} + D^{\alpha}_{\mu \gamma} \tilde{\Gamma}^{\mu}_{\beta 
\delta} \nonumber \\ 
+ \tilde{\Gamma}^{\alpha}_{\mu \gamma} D^{\mu}_{\beta \delta} - D^{\alpha}_{\mu
\delta} \tilde{\Gamma}^{\mu}_{\beta \gamma} - \tilde{\Gamma}^{\alpha}_{\mu 
\delta} D^{\mu}_{\beta \gamma} , 
$$
which leads, after some rearrangements, to
\begin{equation}
\tilde{R}^{\alpha} \:_{\beta \gamma \delta} = R^{\alpha} \:_{\beta \gamma 
\delta} + \tilde{\nabla}_{\delta} D^{\alpha}_{\beta \gamma} - \tilde{\nabla}_{
\gamma} D^{\alpha}_{\beta \delta} + D^{\alpha}_{\mu \delta} D^{\mu}_{\beta 
\gamma} - D^{\alpha}_{\mu \gamma} D^{\mu}_{\beta \delta} 
\label{eq:4}
\end{equation}
Recalling equation (\ref{eq:D2}) we end up with
$$
\tilde{\nabla}_{\delta} D^{\alpha}_{\beta \gamma} - \tilde{\nabla}_{\gamma} 
D^{\alpha}_{\beta \delta} = \frac{1}{2} [ \delta^{\alpha}_{\gamma} \tilde{
\nabla}_{\delta} A_{\beta} - \delta^{\alpha}_{\delta} \tilde{\nabla}_{\gamma} 
A_{\beta} + \tilde{g}_{\beta \delta}\tilde{\nabla}_{\gamma} A^{\alpha} - 
\tilde{g}_{\beta \gamma}\tilde{\nabla}_{\delta} A^{\alpha} ] 
$$
$$
D^{\alpha}_{\mu \delta} D^{\mu}_{\beta \gamma} - D^{\alpha}_{\mu \gamma} D^{\mu
}_{\beta \delta} = \frac{1}{4} [ \delta^{\alpha}_{\delta} A_{\beta} A_{\gamma} 
- \delta^{\alpha}_{\gamma} A_{\beta} A_{\delta} + \delta^{\alpha}_{\gamma} 
\tilde{g}_{\beta \delta} A_{\mu} A^{\mu} \nonumber \\ - \delta^{\alpha}_{\delta
} \tilde{g}_{\beta \gamma} A_{\mu} A^{\mu}  + \tilde{g}_{\beta \gamma}  A^{
\alpha} A_{\delta} - \tilde{g}_{\beta \gamma}  A^{\alpha} A_{\delta} ] 
$$
Introducing these expressions into (\ref{eq:4}), we find
\begin{eqnarray}
\tilde{R}^{\alpha} \:_{\beta \gamma \delta} = R^{\alpha} \:_{\beta \gamma 
\delta} + \frac{1}{2} [ \delta^{\alpha}_{\gamma} \tilde{\nabla}_{\delta} A_{
\beta} - \delta^{\alpha}_{\delta} \tilde{\nabla}_{\gamma} A_{\beta} + \tilde{g
}_{\beta \delta}\tilde{\nabla}_{\gamma} A^{\alpha} - \tilde{g}_{\beta \gamma}
\tilde{\nabla}_{\delta} A^{\alpha} ] +\nonumber \\
\frac{1}{4} [ \delta^{\alpha}_{\delta} 
A_{\beta} A_{\gamma} - \delta^{\alpha}_{\gamma} A_{\beta} A_{\delta} + \delta^{\alpha}_{\gamma} 
\tilde{g}_{\beta \delta} A_{\mu} A^{\mu} - \delta^{\alpha}_{\delta} \tilde{g}_{
\beta \gamma} A_{\mu} A^{\mu}  + \tilde{g}_{\beta \gamma}  A^{\alpha} A_{
\delta} - \tilde{g}_{\beta \gamma}  A^{\alpha} A_{\delta} ]
\end{eqnarray}

We now define the physical and dynamical Ricci tensors in the usual way by
$\tilde{R}_{\beta \delta} \equiv \tilde{R}^{\alpha} \:_{\beta \alpha \delta}$
and $R_{\beta \delta} \equiv R^{\alpha} \:_{\beta \alpha \delta}$, and 
obtain, after some algebra:
$$
\tilde{R}_{\mu \nu} = R_{\mu \nu} + \tilde{\nabla}_{\nu} A_{\mu} + \frac{1}{2}
\tg \tilde{\nabla}^{\alpha} A_{\alpha} - \frac{1}{2} A_{\mu} A_{\nu} + 
\frac{1}{2} \tg A_{\alpha}A^{\alpha} 
$$
Recalling the definition (\ref{eq:A}) of $A_{\alpha}$,  we then calculate
$$
\tilde{\nabla}_{\nu} A_{\mu} = \frac{\tilde{\nabla}_{\nu}\tilde{\nabla}_{\mu}
\F}{\F} - \frac{\F_{, \mu} \F_{, \nu}}{\F^2}, \qquad
\tilde{\nabla}^{\alpha} A_{\alpha} = \frac{\Box_{\tilde{g}} \F}{\F} - 
\frac{\F_{, \alpha} \F^{, \alpha}}{\F^2},
$$
which, when substituted in the above relation between the Ricci tensors, give:
\begin{equation}
\tilde{R}_{\mu \nu} = R_{\mu \nu} + \frac{1}{\F} \left( \tilde{\nabla_{\mu}} 
\tilde{\nabla_{\nu}} \F + \frac{1}{2}\tg \Box_{\tilde{g}} \F \right) -
\frac{3}{2} \frac{\F_{,\mu} \F_{,\nu}}{\F^2} 
\end{equation}

This is the desired relation (\ref{eq:3.4}) we set out to prove.

\end{document}